\documentclass{jkas}

\def\year{2016} 
\def\month{December} 
\def\volume{49} 
\def\beginpage{00} 
\def\endpage{00} 
\def\received{---} 
\def\accepted{---} 
\date{Received \received; accepted \accepted}

\usepackage{balance}
\usepackage{CJKutf8}
\usepackage{verbatim}
\usepackage{amsmath}
\usepackage{amsfonts}
\usepackage{microtype}
\usepackage{subfig}

\def\simlt{\lower.5ex\hbox{$\; \buildrel < \over \sim \;$}}
\def\simgt{\lower.5ex\hbox{$\; \buildrel > \over \sim \;$}}
\def\kms{km\,s$^{-1}$}

\def\msun{{$M_\odot$}}

\def\mguest{m_{V,\,{\rm guest}}}
\def\mimpsoter{M_{V,\,{\rm impostor}}}

\title{The Korean 1592--1593 Record of a Guest Star:\\ An `impostor' of the Cassiopeia A Supernova?}
\author[1]{Changbom~Park}
\author[2]{Sung-Chul~Yoon}
\author[2,3]{Bon-Chul~Koo}
\affil[1]{Korea Institute for Advanced Study, 85 Hoegi-ro, Dongdaemun-gu, Seoul 02455, Korea; \email{cbp@kias.re.kr}}
\affil[2]{Department of Physics and Astronomy, Seoul National University, Gwanak-gu, Seoul 08826, Korea
\email{yoon@astro.snu.ac.kr, koo@astro.snu.ac.kr}}
\affil[3]{Visiting Professor, Korea Institute for Advanced Study, Dongdaemun-gu, Seoul 02455, Korea}
\def\corrauthor{B.-C. Koo}
\def\runningauthor{Park et al.}
\def\runningtitle{The Korean 1592--1593 Record of a Guest Star}
\def\keywords{history and philosophy of astronomy --- supernovae: individual (Cassiopeia A)}
\def\abstracttext{
The missing historical record of the Cassiopeia A (Cas~A) supernova (SN) event 
implies a large extinction to the SN, 
possibly greater than the interstellar extinction to the current SN remnant. 
Here we investigate the possibility that the guest star that appeared near Cas~A 
in 1592--1593 in Korean history books 
could have been an `impostor' of the Cas~A SN, 
i.e., a luminous transient that appeared to be a SN but did not destroy 
the progenitor star, with strong mass loss 
to have provided extra circumstellar extinction.
We first review the Korean records and show that a spatial coincidence between the 
guest star and Cas~A cannot be ruled out, as opposed to previous studies.
Based on modern astrophysical findings on core-collapse SN, 
we argue that Cas~A could have had an impostor and 
derive its anticipated properties.
It turned out that 
the Cas~A SN impostor must have been bright 
($M_V =-14.7 \pm 2.2$~mag) and 
an amount of dust with visual extinction of $\ge 2.8\pm 2.2 $ mag
should have formed in the ejected envelope and/or in a strong wind afterwards.
The mass loss needs to have been spherically asymmetric 
in order to see the light echo from the SN event but not the one from the impostor event.
}

\begin{document}
\begin{CJK}{UTF8}{bsmi} 

\jkashead 


\section{Introduction}

During the last thousand years, five supernova (SN) events in the Milky Way 
have been witnessed and recorded in history books 
of East Asian and/or European countries: SN 1006, 1054, 1181, 1572, and 1604
\citep{ste02}.
Each of them was visible to the naked eye in the night sky over a period longer than six months,  
and their supernova remnants (SNRs) 
are now observable as beautiful nebulae with modern telescopes.
The SN event that produced the SNR Cassiopeia A (hereafter Cas~A), 
however, was an exception.

Cas~A is a small, spherical remnant expanding rapidly 
($\sim 5\,000$~\kms) at a distance 3.4 kpc (see Figure 1).
It was discovered in 1940s as a bright radio source \citep{ryle1948}, 
and since then it has been extensively studied over all wavebands.
In particular, the SN flash light at the time of explosion was 
detected in 2008 as the SN light ‘echo’, 
which confirmed that Cas~A is a remnant of core-collapse SN (CCSN) with a massive 
(15--20~\msun) progenitor, i.e., SN IIb \citep{kra08}. 
The material that we see 
in Figure 1 is mostly heavy elements 
synthesized in the innermost region of the progenitor 
star and expelled by the explosion.
The proper motion studies of almost freely-expanding 
dense SN material have yielded 
an accurate date of the SN event, i.e., AD $1671.3 \pm 0.9$ and $1680.5\pm 18.7$ 
\citep{tho01,fes06a}, implying that 
the Cas~A SN event was in the telescope era. 

\begin{figure}[b!]
\centering
\includegraphics[trim=5mm 22mm 5mm 10mm, clip, width=84mm]{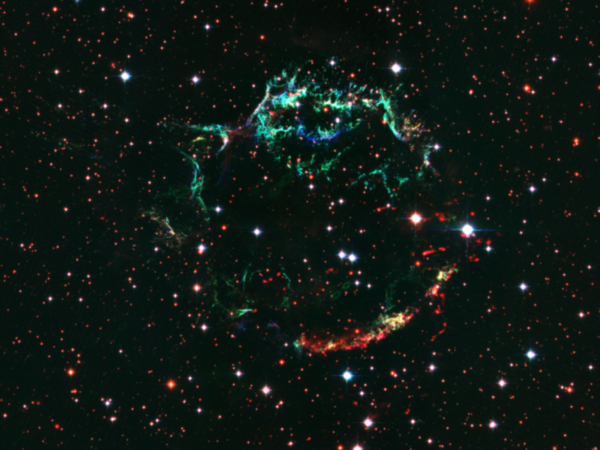}
\caption{
Three-color composite optical/near-infrared image of the Cas~A SNR 
produced from HST ACS F775W and F850LP (B and G) and Palomar WIRC
[Fe II] 1.64 $\mu$m (R) image \citep{ham08,lee15}. The emission in HST ACS F775W and F850LP filters
are dominated by [O II] and [S III] emission lines, respectively.
}
\label{fig:1}       
\end{figure}

It is, therefore, surprising that 
there is no uncontroversial historical record on the Cas~A SN event.
It has been suggested that 
the 6-th magnitude star `3~Cas' recorded by 
John Flamsteed on August 16 in 1680 corresponds to the Cas~A SN event 
\citep{ash80}. But this object is most likely a non-existing star resulting from   
combining measurements of two different stars by mistake \citep{bro79,ste02}.
An extensive 
search using the historical 
astronomical records of three East Asian countries has also 
failed to find any record matching the Cas~A SN event \citep[e.g.,][]{ste02}.
The only records of a guest star that appeared very close to Cas~A was 
made by Korean astronomers in 1592--1593, for which  
\citet{bro67} and \citet{chu68} independently 
proposed that it might have been the Cas~A SN event. 
On the other hand, \citet{ste87} \citep[see also][for a summary]{ste02}, who studied this Korean record in detail, concluded that not only the 
date but also the positional agreement between the guest star 
and Cas~A was poor, 
so that this Korean record was attributed to a nova. 

Modern astrophysical studies suggest that the Cas~A SN 
might have been a normal 
CCSN with absolute magnitude of about $-17.5$ \citep[][see also Section~4]{you06,kra08}, so that 
at a distance of 3.4 kpc, it should have been visible 
if the interstellar extinction were `typical'.  
The extinction to Cas~A is indeed large ($A_V=4$--8 mag), which may or may not 
be enough to explain the missing historical 
records \citep[][see also Section~4]{koo16}. 
An interesting hypothesis is that there was  an 
extra extinction at the time of SN explosion, 
possibly due to mass loss from the progenitor star \citep{pre95,har97}. 
This raises the question if 
the guest star in 1592--1593 could have been such a mass loss event of Cas~A.
It is known that some CCSNe experience luminous transients, sometimes as bright as $M_B=-17.8$~mag  \citep{koc12}, 
and ejection of mass without destruction of the stars. 
Such `SN impostor' phenomena have been observed mostly in SNe of type IIn, but their nature is poorly known.
A significant fraction of SN impostor progenitors are indeed relatively low mass (15--25~\msun) stars 
\citep[][and references therein]{koc12,ada15}.
Therefore, although the reported years are too early to be directly responsible for Cas~A, there could be a possibility 
that the Korean records of the guest star in 1592--1593 
was an impostor that provided extra circumstellar extinction to the 
Cas~A SN in 1670s. 
In Section~2, we review the Korean records and show that the spatial coincidence between the 
guest star and Cas~A cannot be ruled out, as opposed to previous studies.
In Section~3, we discuss 
the possibility that the guest star was an impostor of the Cas~A SN 
based on modern astrophysical findings on CCSN. And in Section~4, we derive 
the properties of the Cas~A SN impostor that can explain the observations and conclude our paper. 

\section{Korean Records of the 1592-1593\\ Guest Star}

Korean astronomical records during {\it Joseon} dynasty (1392--1910) are to be found in the official history books of Joseon dynasty: 
{\it Joseon Wangjo Sillok} 朝鮮王朝實錄 (the Annals of the {\it Joseon} Dynasty) and 
{\it Jeungbo Munheon Bigo} 增補文獻備考(Enlarged Official Encyclopedia). 
In addition to them, there are other historical books that contain astronomical records, such as 
{\it Seungjeongwon Ilgi} 承政院日記 (Diary of the Royal Secretariat) and {\it Ilseongnok} 日省錄 (Cords of Daily Reflection). The contents of these books are available in the Internet (\url{http://db.history.go.kr} and \url{http://www.minchu.or.kr}). We inspected all these books to search for records that can match the Cas A event. The only records possibly associated with Cas A were the sightings of a guest star near Cas A in 1592-1593 
 in the Annals of the {\it Joseon} Dynasty that had been already 
reported in previous studies \citep{bro67, chu68, ste02}. In the following, we discuss these records. 

\begin{figure}[t!]
\centering
\includegraphics[trim=0mm 7mm 0mm 12mm, clip, width=84mm]{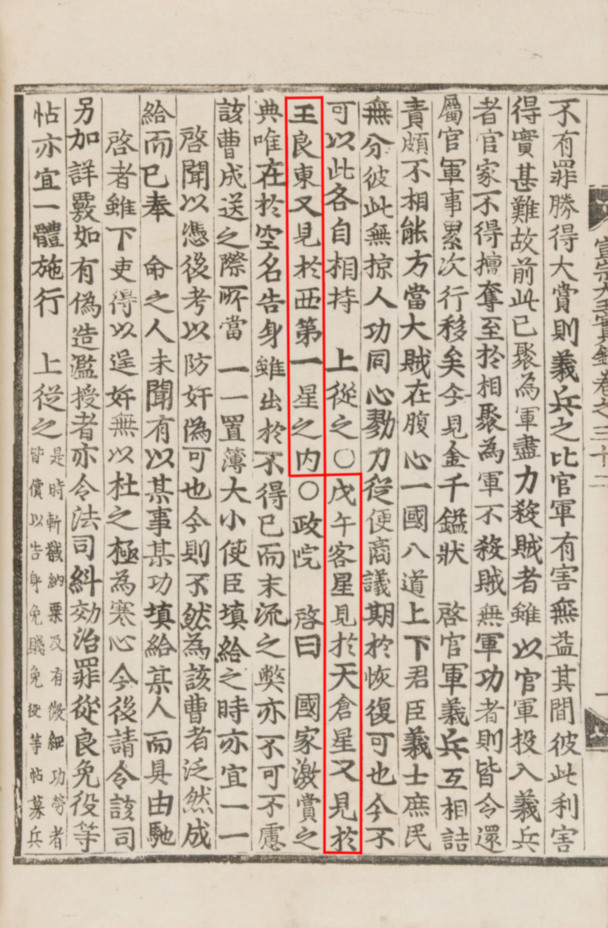}
\caption{
A page in  {\it Seonjo Sillok} of {\it Joseon Wangjo Sillok} 朝鮮王朝實錄 (the Annals of the {\it Joseon} Dynasty) showing the record of sighting of a guest star that appeared inside (內) the first star ($\beta$ Cas) of the constellation {\it Wangyang} 王良  on December 5, 1592. This is the second record of the guest star in the book after the one a day before.}  
\label{fig:2}       
\end{figure}

\begin{figure*}[t!]
\centering
\subfloat{\includegraphics[width=64mm]{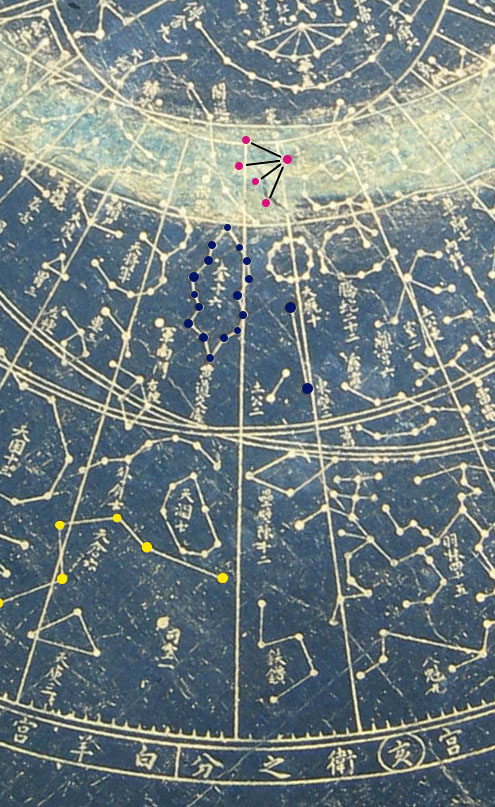}}
\hspace{7em}
\subfloat{\includegraphics[width=70mm]{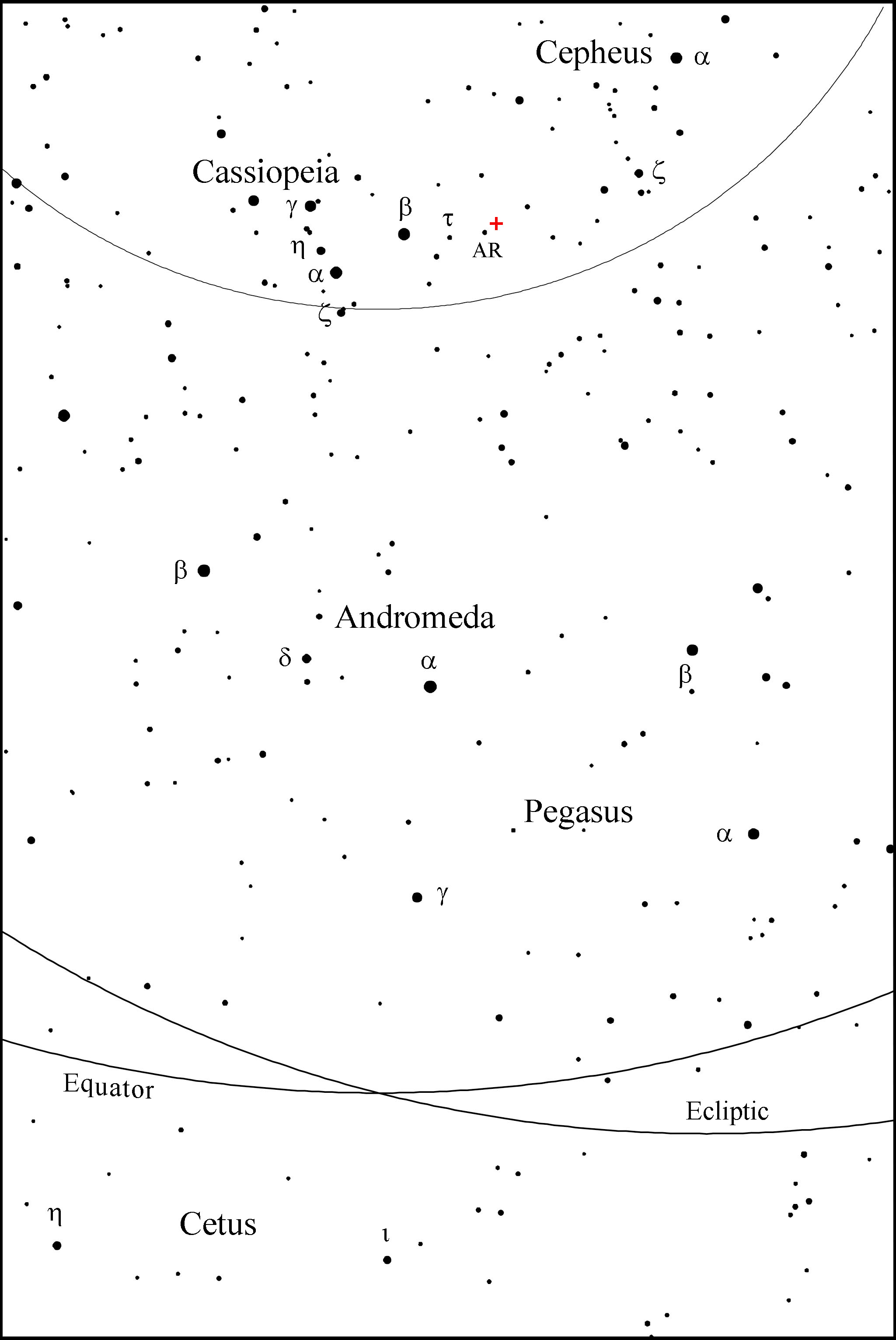}}
\caption{\emph{Left:} A part of {\it Cheonsang Yeolcha Bunyajido} 
天象列次分野之圖, 
the old Korean constellation map of {\it Joseon} dynasty made in 1395, 
showing the constellations {\it Wangyang} 王良, pink dots), 
{\it Byeok} (壁, right two dark blue points), 
{\it Gyu} (奎, left 16 dark blue points), 
and {\it Cheonchang} (天倉, yellow dots at bottom left). 
The right most star of {\it Wangyang} is $\beta$ Cas, and the right-most star 
of {\it Cheonchang} is $\iota$ Cet. The epoch of this map is around 1st century A.D (Park 1998).
\emph{Right:} Position of the stars near Cassiopeia and Cetus brighter than 5.5 in visual magnitude in 1592. 
The expansion center of Cas~A is marked with a small cross near AR Cassiopeia. 
The angular distance between $\beta$ Cas and the expansion center is 5.9 degrees.} 
\label{fig:3}       
\end{figure*}

The first record appears on December 4, 1592 (November 1 in lunisolar calendar) in {\it Seonjo Sillok} 宣祖實錄  of the Annals of the {\it Joseon} Dynasty, which reports that a guest star appeared ``at the first star in the west of {\it Wangyang} 王良''. After this first record the location of the guest star was consistently described as “inside (內) the first star in the west of {\it Wangyang}” (Fig. 2). The total duration of the sightings of the guest star in the record is 91 days. The west-most star in the constellation of {\it Wangyang} is $\beta$ Cassiopeia ($\beta$ Cas). 
The angular distance between $\beta$ Cas and the expansion center of Cas A determined from the proper motion of SN ejecta knots \citep{tho01} is 5.9 degrees. 

\cite{ste87} \citep[see also][]{ste02} studied these Korean records in detail, and concluded that these guest star records were probably observations of a nova rather than sightings of Cas A supernova. \cite{ste87} tried to find the meaning of “inside (內)” by examining the distance between objects whose separation was described with the same expression. They noted that there are records of another guest star appearing in the constellation {\it Cheonchang} 天倉 from November 23, 1592 to September 15, 1594, which describe the position of the guest star to be “inside the third star at the east of {\it Cheonchang} and about 3 {\it chon} 寸 away''.
They estimated 3 {\it chon} to be about 0.45 degree and concluded that the expression “inside (內)” was used when a separation was much less than one degree. This led them to conclude that the separation of 5.9 degrees between $\beta$ Cas and Cas A was too large to be described by the expression that “the guest star was inside the star $\beta$ Cas”.

Since the meaning of “inside (內)” was critically important in their rejection of the records of the guest star near $\beta$ Cas as the sightings of Cas A SN, 
we investigated other usages of the same expression in the same history book, 
{\it Seonjo Sillok}. We found another record on October 21, 1585, 7 years before the above records, saying that “A comet appeared inside the first star at the west of {\it Cheonchang}. It was separated by 7 degrees from  {\it Byeokseong} 壁星 and by 101.5 degrees from the north pole star.” Here {\it Byeokseong} and the first star at the west of {\it Cheonchang} correspond to 
$\gamma$ Peg and $\iota$ Cet, respectively \citep{park1998}.  
Note that, in this record, the angular separation of 7 degrees is in right ascension and that the angles were still measured in the traditional system where the angle of a full circle was 365.25 degrees instead of 360 degrees. Precession calculation of $\gamma$ Peg and $\iota$ Cet to the year of 1585 gives angular distance between this comet and $\iota$ Cet of 5.4 degrees, which is much larger than one degree.

In the above calculation, however, we find that the comet would be located within the lunar mansion of {\it Gyu} 奎 instead of {\it Byeok} 壁. If this were the case, the record would have reckoned angular separation from 
the constellation {\it Gyu} instead of {\it Byeok}. 
A possible explanation would be that they measured the right ascension of the comet from the old royal Korean constellation map, {\it Cheonsang Yeolcha Bunyajido} 天象列次分野之圖, presented in Figure 3 (left panel). In this map, $\iota$ Cet is 1~degree 
east of the boundary between the {\it Gyu} and {\it Byeok} lunar mansions 
and since the comet was 7 degrees from 
{\it Byeok} (or 2 degrees west of the {\it Gyu}-{\it Byeok} lunar mansion boundary) according to the record, the difference between the comet and $\iota$ Cet in right ascension is about 3 degrees in the traditional angle-measuring system. 
If the north polar distance is used as recorded, the total angular separation between the comet and $\iota$ Cet is 3.1~degrees.
In either case, the separation intended by the expression “inside a star” was much larger than one degree, as large as 3.1 or 5.4 degrees. 
\cite{ste87}’s understanding of the expression was only based on one record for which the separation happened to be less than one degree. The fact that the expression “inside a star” was used for separations up to a few degrees allows one to keep the records of the guest star near $\beta$ Cas in 1592--1593 as an event possibly associated with Cas A.

Before closing this section, it is worthwhile to comment on one false guest star and two other guest stars recorded in {\it Seonjo Sillok} during the same period. 
The false one is the above mentioned guest star in {\it Cheonchang}, for which \cite{ste87} studied the meaning of “inside”. On September 15, 1594, however,  
the court astronomers requested that its identification as a guest star be canceled after reporting to King {\it Seonjo} that the guest star must be a fixed star because it had been visible for more than three years. 
This implies that they made self-inspection of their reports on guest stars. The second one is the guest star between the first and second eastern stars of {\it Wangyang}, which appeared from November 30, 1592 to March 28, 1593 for 119 days. 
This may be a nova event as argued by \citet{ste02}. There are only two mentions of the third guest star 
separated by  37 days first appearing on December 12, 1592 above the constellation {\it Gyuseong} 奎星. 
This could have been a nova event too as suggested by \citet{ste02}.

\section{impostor of the Cas A Supernova?}

The discussion in the previous section suggests that the guest star in the Korea
1592--1593 records could be spatially coincident with the Cas A SN. The
observation date of this guest star, however, is $\sim$80 years before the
estimated date of the Cas A SN, which makes the Korean records in 1592--1593 unlikely
to be the sightings of the SN event responsible for the Cas A SNR observed
today. 

We may still consider the possibility that the guest star was related to a
strong eruption of the progenitor star of the Cas A SN.  It is well
known that some massive stars  undergo an outburst of the envelope during the
course of their evolution, which would appear as a supernova impostor (see
Smartt 2009, Kochanek et al. 2012 for a review).  A notable example is the Great 
Eruption of $\eta$ Carinae that occurred in the 19th century. 
Most of SN
impostors appear less bright than a supernova, typically having visual
magnitudes of $-15 \lesssim M_V \lesssim -11$.  The duration of their optical
transients ranges from about 20 days to 4400 days. 

Steady, line-driven winds cannot explain this phenomenon and the physical
mechanisms of these outbursts have not been well understood yet.  For very
massive stars close to the Eddington limit, various hydrodynamic instabilities
like the strange-mode instability may occur in the envelope, which can trigger
non-steady mass outflows on a short timescale (Owocki 2015). This may explain
eruptions from luminous blue variables (LBVs).  However, not all SN impostors
would originate from very massive stars.  Binary interactions that  lead to a
common envelope ejection or a merger may also produce an event that resembles a
SN impostor (e.g., Ivanova et al. 2013).  Nuclear shell
flashes during the late stages of  massive star evolution may also trigger an
outburst of the outer envelope from a massive star (Woosley \& Heger 2015).
Binary interactions do not necessarily involve very massive stars, and the
latter mechanism strongly favors the low-mass end of CC progenitors
($9 - 11~M_\odot$).

Observations indicate that a strong outburst of the envelope can occur shortly
before the real SN explosion. For example, SN 2009ip was a SN impostor from a
LBV, and turned into a true SN three years later, which appeared as a SN~IIn
(Mauerhan et al. 2013).  Could the guest star in 1592--1593 have been a SN
impostor as a precursor of the Cas A SN as well? The time span of about 80
years between the appearance of this guest star and the Cas~A~SN is very long
compared to the case of SN~2009ip for which the real SN explosion in 2012
occurred 3 yrs later than its precursor. However, such an outburst can occur at
anytime before the SN explosion in principle, in particular if binary
interactions are responsible for it.  

The case of SN~2014C is particularly relevant in this regard. The recent study
on SN~2014C, which appeared as a SN~Ib initially and evolved into a SN~IIn over
one year, indicates that the progenitor  experienced a mass eruption about  100
years before the SN explosion (Milisavljevic et al. 2015; Margutti et al.
2016).  A situation similar to this case may have occurred with Cas A.  The
fact that the Cas A SN is type IIb instead of IIn does not necessarily
contradict this scenario. 
The light echo spectra of the Case A SN by \cite{kra08} are believed to reflect 
rather an early stage of the SN, i.e.,  near the optical peak.
We cannot exclude the possibility
that the Cas~A~SN would have evolved into type IIn in later stages as in the
case of SN 2014C.  Alternatively, it would not have turned into a SN IIn if the
mass of the ejected material during the outburst in 1592 had been sufficiently
small.  This issue deserves future work with numerical simulations. 

\section{Discussion and Conclusion}
  
In this section, we discuss the properties of the Cas A SN impostor 
anticipated if the guest star in the Korean 1592--1593 records had been 
an impostor of the Cas A SN and had provided 
necessary extra extinction to hide the SN event.

We first consider the brightness of the impostor.
There is no record about the brightness of the guest 
star.\footnote{\cite{ste87} in their paper said 
``The only brightness estimate for 1592C is that it rivalled 
Praesepe (i.e., 5th mag) soon after discovery.'' 
where 1592C is the guest star in this paper, but we could not confirm this. 
There is a record on November 4, 1592, i.e., 
four days after the discovery of the guest star, 
on the brightness of {\it Jeoksisung} 積尸星 which is Praesepe in modern constellation, 
but it is a separate observation irrelvant to the guest star. }
Considering that the stars in the old Korean constellation map 
{\it Cheonsang Yeolcha Bunyajido} 
are mostly brighter than 5 mag, however, 
the guest star might have been brighter than $\sim 5$ mag to be witnessed  
as a `guest' star. 
On the other hand, it was probably not very bright ($\ge 3$~mag) 
because otherwise 
its brightness would have been compared to the known 
bright stars or planets, e.g., the brightness of the 
Kepler SN on April 23, 1605 was compared to the 2.8 mag star $\tau$ Scopi 
in {\it Seonjo Sillok}. 
Therefore, we may assume that the brightness of the guest star was 
$\mguest=4 \pm 1$~~mag.
At the distance of Cas A (3.4 kpc), which implies that the absolute magnitude of 
the impostor should be  
\begin{align}
\mimpsoter &=\mguest - 5 \log d+5-A_V \nonumber \\
                    &=-8.7\pm 1.0 - A_V,
\end{align}
where $A_V$ is the extinction to the impostor. 
There have been considerable efforts to estimate the extinction to 
the explosion center of Cas A \citep{tro85, keo96, rey02, eri09}, and excluding  
the results from X-ray studies (see below), 
the derived extinction ranges 6--8 mag (see also Table 3 of \citealt{koo16}).
These numerical values, however, are partly on the high side, and, 
considering various uncertainties in converting the observed physical parameters 
to the extinction, we adopt $A_{V,{\rm ISM}} = 6\pm 2$ mag for the  
{\em interstellar} extinction to the expansion center of Cas A.
%
%
Therefore, assuming that the extinction to the impostor in Equation (1) is  
given by $A_{V,{\rm ISM}}$,   
we obtain $\mimpsoter =-14.7 \pm 2.2$~mag, which is comparable to the 
brightest SN impostors.   

We next consider the magnitude of the circumstellar extinction to be 
provided by the impostor ($\Delta A_{V,\,{\rm circum}}$) for the Cas A SN event to be 
missed by `contemporary' observers.
The peak brightness of a hydrogen-deficient SN is directly proportional to the mass of synthesized $^{56}{\rm Ni}$ mass \citep{arn82}.
For Cas A, the total $^{56}{\rm Ni}$ mass inferred from the observed $^{56}{\rm Fe}$ and  $^{44}{\rm Ti}$ masses 
is $\sim 0.2$~\msun \citep{hwa12, you06}, 
which is comparable to the average $^{56}{\rm Ni}$ mass of SN Ib \citep{dro11}, 
so that the peak brightness of Cas A is believed to have been 
close to the mean $M_V$ ($=-17.6\pm 0.9$ mag) of SN Ib.  
We adopt $M_{V,{\rm SN}}=-17.5$ mag, which was the peak brightness of SN 1993J, 
as the characteristic absolute brightness of the Cas A SN.
If we assume that the apparent peak brightness of SN was fainter than $4$ mag 
to be missed by the 17th-century observers \citep[e.g.][]{shk68}, although 
we cannot exclude the possibility that the 
SN was brighter but overlooked (see Koo \& Park 2016 for a summary), 
we obtain 
\begin{align}
\Delta A_{V,\,{\rm circum}} &=m_{V,{\rm SN}}-M_{V,{\rm SN}}-5\log d+5-A_V\nonumber \\
                                       &\ge 2.8\pm 2.2~{\rm mag}.
 \end{align}                                 
This appears to be a large extinction for an expanding dusty shell,  
although $\eta$ Carina might have had produced a large amount of dust 
in its Great Eruption (1837--1856)  
which made its second eruption in 1890s fainter by $\sim 4.3$~mag \citep{hum99}.
The outburst, however, is thought to be a signal that the star is entering 
a high-mass loss rate phase for some SN impostors \citep{koc12,ada15}, 
and the Cas A progenitor, survived from the eruption,  
could have had a strong, steady dusty wind. 

In conclusion, in order for the guest star 
in the Korean 1592--1593 historical records 
to be an impostor of the Cas A SN  
that had provided extra circumstellar extinction to hide the SN event,
the impostor must have been bright  
($\mimpsoter =-14.7 \pm 2.2$~mag) and 
an amount of dust with $\Delta A_{V,\,{\rm circum}}\ge 2.8\pm 2.2 $ mag
should have formed in the ejected envelope and/or 
in the strong wind afterwards. 
It is not likely that we can see a footprint of such mass loss in the 
$\sim 340$~yr-old SNR, 
although X-ray studies suggested   
a cavity of 0.2--0.3 pc radius which was attributed to 
fast Wolf-Rayet wind \citep{hwa09}.
The larger extinction implied from 
X-ray absorbing gas columns, i.e., 9--11 mag \citep{wil02,hwa12} 
has been suggested as evidence for  
the circumstellar gas where dust grains had been present but 
destroyed by SN shock, but the extinction measurements are uncertain
and the evidence is circumstantial.

In the impostor hypothesis, it is likely that the impostor in 1592 AD 
was brighter than the Cas A SN in the 1670s in apparent magnitude.  
This raises a question: do we or can we see the light echo of the impostor? 
The spectra of infrared (IR) light echos of Cas A SN have been shown to be 
consistent with thermal emission from interstellar dust heated by 
an intense and short burst EUV-UV radiation from the SN shock breakout \citep{dwek08}.
On the other hand, the optical spectra of the light echo obtained at three different positions 
all look similar and consistent with SN IIb \citep{rest11}.
Therefore, all the detected IR/optical light echos appear to be 
associated with the SN event, and there is no evidence for the 
light echo of the impostor event.
It is not impossible that there is one to be detected, but
perhaps a more plausible conjecture is that 
the mass loss of the impostor was not spherically symmetric 
so that its circumstellar extinction affected only some limited sight lines, e.g., the direction toward the Earth.      
If such were the 
case, the Cas A SN event will be much (by $\sim 3$ mag) brighter than the impostor event toward most directions, and we may only see the echo of the SN event. 
Considering that the jet axis in the Cas A SNR is almost perpendicular to the line-of-sight,
if the mass loss was preferentially along the equatorial plane that happens to coincide with 
our line-of-sight, this does not seem to be unplausible.

\acknowledgments

We wish to thank Jae-Joon Lee, Dave Green, and Dan Milisavljevic 
for their helpful comments on the manuscript. We also thank the anonymous referee 
for his/her comments on the presentation of Korean historical records.
We gratefully acknowledge the support from the JKAS Editorial Office on using Chinese characters in
the paper and the latex help by Sascha Trippe.
BCK was supported by Basic Science Research Program through the National Research
Foundation of Korea (NRF) funded by the Ministry of Science, ICT and Future Planning
(2014R1A2A2A01002811).
%


\end{CJK} 
\end{document}